# Rise of QAnon:
# A Mental Model of Good and Evil Stews in an Echochamber

**J. Hunter Priniski**
(priniski@ucla.edu)

**Mason McClay**
(masongmcclay@ucla.edu)

**Keith J. Holyoak**
(holyoak@psych.ucla.edu)

Department of Psychology
University of California, Los Angeles

**Abstract**

The QAnon conspiracy posits that Satan-worshiping Democrats operate a covert child sex-trafficking operation, which Donald Trump is destined to expose and annihilate. Emblematic of the ease with which political misconceptions can spread through social media, QAnon originated in late 2017 and rapidly grew to shape the political beliefs of millions. To illuminate the process by which a conspiracy theory spreads, we report two computational studies examining the social network structure and semantic content of tweets produced by users central to the early QAnon network on Twitter. Using data mined in the summer of 2018, we examined over 800,000 tweets about QAnon made by about 100,000 users. The majority of users disseminated rather than produced information, serving to create an online echochamber. Users appeared to hold a simplistic mental model in which political events are viewed as a struggle between antithetical forces—both observed and unobserved—of Good and Evil.

## Introduction

Deception and misinformation have pervaded American politics throughout its history. During the Civil War, Southern sentiment was manipulated by press campaigns propagating unfounded claims that the North was planning to lead a revolt of freed slaves against the South (Necklason, 2020). A century and a half later, false allegations of stockpiled weapons of mass destruction in Iraq served to justify a U.S. invasion. This pretense remained a pervasive belief among the American public for at least a decade (Altheide & Grimes, 2005). More recently, online forums promoted a conspiracy theory asserting that Jewish financial elites helped orchestrate the Covid-19 pandemic (Zipperstein, 2020). The emergence of social media has enabled unregulated production and proliferation of misinformation—a global "infodemic" of misinformation (Salvi et al., 2020)—making our era a Golden Age of conspiracy theories.

One of the most widespread fantasies to emerge from recent internet forums is the QAnon conspiracy (Zipperstein, 2020; Papasavva et al., 2020; Saltman, 2020). QAnon's central narrative involves a covert world of liberal elites ruling by way of a global child-trafficking ring and by mass manipulation of the economy and media. As the main protagonist of QAnon, former U.S. President Donald Trump is purported to be the sole defense against this elite cabal. While this narrative is disturbing and radical, the most nefarious aspects of QAnon are likely its varied peripheral narratives and claims that sow distrust in scientific and democratic institutions (Zipperstein, 2020). For instance, during the beginning of the Covid-19 pandemic, a meme posted on a QAnon 4Chan board falsely claimed that 96% of reported deaths attributed to Covid-19 were not actually due to that disease. While this claim was easily refutable, within days it inspired massive anti-lockdown protests in Germany, the United Kingdom, and the United States, which were primarily attended by protestors holding signs that bore the QAnon slogan, *WWG1WGA* ("where we go one, we go all").

Belief in QAnon's web of narratives is not reserved for those on the political extremes. A CIVIQS (2020) survey reported that one in three Republicans believe that QAnon is mostly true. Given that a substantial proportion of the electorate is comfortable with a radically inaccurate depiction of reality, the pervasiveness of QAnon-supporting sentiment may pose a threat to the stability of Western democratic institutions (Saltman, 2020). Understanding the social and cognitive factors enabling the rise of QAnon is crucial to prevent future emergence of similar conspiracies.

## Hypercoherence as a Prerequisite for Conspiratorial Thinking

Generalization of previously learned information to novel situations is a hallmark of adaptive learning (Mednick & Freedman, 1960). In the context of belief formation, new beliefs also tend to generalize from, or cohere with, features of prior beliefs (Lewandowsky, Gignac, & Oberauer, 2013; Homer-Dixon et al., 2013). Indeed, a coherence mechanism has been shown to be central to various cognitive processes, from visual perception (Yuille & Grzywacz, 1988) to moral reasoning (Holyoak & Powell, 2016). According to explanatory coherence theory, beliefs are often formed on the basis of congruence with prior beliefs, insofar as the acceptance of a new belief increases the explanatory coherence of the belief network (Findlay & Thagard, 2011). Beliefs may therefore be adopted if they fit the explanatory model generated from one's prior beliefs.

But what happens when new evidence can be construed as coherent with one's prior beliefs, regardless of its veridicality? Recent work has shown that people who engage in conspiratorial thinking tend to attribute more control and structure to the world than is plausible. Conspiratorial thinkers do not typically believe in just a single conspiracy theory, but rather clusters of them (van Harreveld et al., 2014). For example, Lewandowsky, Oberaurer, and Gignac (2013) found that people's propensity to believe that NASA faked the moon-landing predicted their tendency to believe that climate change was a hoax. This association arises because conspiracy thinkers are likely to endorse completely novel conspiracies that share common conspiracy themes, thus viewing logically disjoined narratives as mutually coherent (e.g., NASA/government conspiracy → fake moon landing; climate scientists conspiracy → climate science is fake). This evidence suggests that conspiracy thinkers may readily bind new information with their conspiratorial view of the world through a maladaptive level of coherence—*hypercoherence*. Hypercoherence combines top-down priors based on broad core attitudes (e.g., distrust of government and scientific elites)*,* coupled with bottom-up "data" based on the opinions of fellow believers that echo on social media. Where conspiracy thinkers go one, they go all by attributing a vast network of complex narratives to a single causal source (Saltman, 2020).

Hypercoherence may be especially easy to achieve when information is consumed within online communities where both information and social identity are radically curated. Conspiracies such as QAnon may be a natural consequence of a social media environment that: (1) prioritizes false information over verifiable information, and (2) allows for the easy and rapid formation of *echochambers*, or pockets of online communities that share and consume nearly identical, belief-confirming information (Sasahara et al., 2020). Once misinformation is introduced that coheres with the narrative of a particular echochamber, it may foster the generation of additional content by simultaneously adding to the coherence of the community's narrative while reducing its standard of plausibility. Misinformation may therefore gradually reconfigure a person's belief network toward stronger degrees of coherence, making it more capable of binding disparate and implausible beliefs. The result is belief in conspiracies that cover a wide range of narrative clusters.

These factors make QAnon no longer merely a single conspiracy, but a web of conspiratorial plots under the umbrage of a central narrative and shared identity (Roose, 2021). Understanding the features of social media networks in which QAnon has spread may help elucidate the conditions under which conspiratorial trends metastasize into super-conspiracies. Such analyses may help to find ways to stymie the growth of future conspiracies that take root in internet discussions. To this end, here we report analyses of the structure and content of the early QAnon Twitter network, aiming to infer the mental model that binds members of this community together while understanding the social processes underlying the narrative's rapid spread online.

## Computational Studies of Twitter Data

We conducted two studies of the early QAnon Twitter network to understand the social and cognitive processes shaping the dissemination and content of the conspiracy's narrative. In Study 1 we analyzed a retweet network to assess the extent to which users shared versus produced content. In Study 2 we fit a series of topic models to the tweets to characterize the general form of the mental model shared by QAnon users.

From June 29 to July 12, 2018, 834787 tweets from 107,777 unique users were collected using the Twitter Streaming API. Tweets containing at least one of the following strings were collected: "qanon", "#q", and "#qanon". In January 2021, after Twitter removed Q accounts following the insurrection on the U.S. Capitol, tweets from many of these accounts were permanently deleted. This early dataset thus provides a rare and crucial glimpse into the echochamber that grew to eventually impact American politics for much of 2020.

## Study 1: Network Analysis

Social networks—both online and in the physical world—are central to the function and maintenance of conspiratorial beliefs. We analyzed a retweet network to examine how different users contributed to the production and distribution of information.

### Retweet Network Construction

Retweet networks capture how information is shared in a Twitter network. Previous work has highlighted the need to identify different user types in order to understand messaging patterns in Twitter data (Kwon, Priniski, & Chanda, 2018). This information may not only help to understand how conspiratorial narratives grow via adoption on social media (i.e., attracting more followers), but also may help identify target points that, if removed, would effectively disrupt the network (e.g., removing a certain user from the platform). A retweet network is represented mathematically as a digraph (directed graph) $G = (V, E)$, where $V$ contains a set of vertices (or nodes) representing unique Twitter users and $E$ contains a set of directed and weighted relations between two users $v_i$ and $v_j$ in $V$, and where $e_{i,j}$ represents the number of times user $v_i$ retweeted $v_j$. Consequently, the *degree* of vertex $v_i$—which equals the sum of all weighted edges connected to $v_i$—is a proxy for the amount of information $v_i$ shared within the network. Because $G$ is a digraph, edges encode directionality and therefore $e_{i,j}$ may not necessarily equal $e_{j,i}$. For instance, the number of times user $v_i$ retweeted $v_j$ may not equal the number of times user $v_j$ retweeted $v_i$. Therefore, a node's degree can be further decomposed into an *out-degree* and *in-degree* component, respectively representing the number of times $v_i$ shared other users' tweets, and how many times other users shared $v_i$'s tweets.

Analysis of a user's in- and out-degree can reveal their information-sharing role in the network. Specifically, high in-degree values indicate the user produced content that was heavily retweeted by other users in the network, suggesting that the person was a *producer* of information in the ecosystem. Producers are likely to be central to the production of information in the echochamber and to development of the central narrative. In contrast, relatively high out-degree values indicate users who more frequently shared other users' content rather than producing their own. Such *distributors* are central to the transmission of misinformation emerging from the core of the echochamber to users outside the network. Distributors may lure new users into the echochamber by exposing them to misinformation via newsfeeds.

Both user types are central to the generation and spreading of political disinformation and synthesis of a unifying narrative (e.g., Keller, Schoch, Styler, & Tang, 2020). Understanding the proportion of users playing each role can reveal how the conspiracy grew and spread. If there is relative balance between the two user types, this would suggest that many users contribute to the narrative by producing information, and the conspiracy is relatively "grass roots" in nature. In contrast, if there are more users with large out-degree values, this would suggest that the majority of users in the network distribute rather than produce information, suggesting that only a minority of highly influential users are key to the production of information in the network.

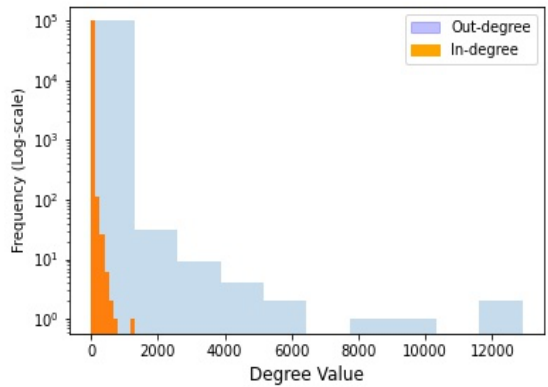

*Figure 1*. Distribution of out-degree (number of times a user retweeted another) and in-degree (number of times a user's original tweet was retweeted by another) values across the QAnon conspiratorial network.

*Table 1.* Summary Statistics Describing Network Structure.

| Network Statistics | |
|---|---|
| Nodes | 98352 |
| Edges | |
|   Unique | 430036 |
|   Weighted Sum | 655216 |
| Out/In-Degree Distribution | |
| Min | 0/0 |
| Max | 12899/1314 |
| Mean | 4.37/4.37 |
| Variance | 9011.5/172.7 |
| Skew | 76.7/23.0 |

As shown in Figure 1, for this dataset the range of out-degrees values is in fact far larger than that of in-degree values, suggesting that most users distributed information produced by a small set of influential users. This interpretation is supported by the differences in statistics describing the distributions of these values (Table 1). Notably, out-degree values were far more variable than in-degree values. Strikingly, the extremes of the degree distributions suggests that some users retweeted up to 13,000 times in the dataset.

## Uncovering Communities and Central Users

Understanding the global organization of a social network can shed light on its operations. Due to the inherent complexity of real-world networks, numerical methods are required to reveal higher-level structures, such as tightly-connected clusters of nodes, or *communities*. Here, we apply the *k-clique* percolation algorithm (or *k*-clique community detection) to uncover overlapping communities across levels of "embeddedness" in the network (Palla, Derenyi, Farkas, & Vicsek, 2005). The goal of this analysis is two-fold: (1) to assess the prevalence of communities across the network, which will guide our understanding of how information is produced and shared across users in the network; (2) to reveal which users are most central to the network in order to analyze their tweet content (Study 2).

*K*-clique community detection finds substructures in a network by first finding all *k*-cliques in a network. A *k*-clique is a set of *k* nodes in a network such that all nodes in the set share an edge between them. In other words, a *k*-clique is a fully connected set of *k* nodes. For example, if there are three nodes, $v_1$, $v_2$, and $v_3$ such that $v_1$ is connected to $v_2$, $v_2$ is connected to $v_3$, and $v_1$ is connected to $v_3$, then the three nodes form a 3-clique. The *k*-clique community detection algorithm uses the sets of *k*-cliques—and the nodes that span multiple separate *k*-cliques—to assess larger substructures in the network. Specifically, a community is constructed between two separate *k*-cliques if at least one node is shared between them. For instance, if nodes $v_1$, $v_2$, and $v_3$ form a 3-clique, and $v_3$, $v_4$, $v_5$, and $v_6$ form a 4-clique, then because the node $v_3$ is common to both cliques, a community of the seven nodes is constructed. By sweeping across values for *k* and extracting communities, we can get a sense of the number of tightly-connected communities at varying sizes. Users belonging to *k*-clique communities with larger values of *k* are more embedded within the network.

The number of communities resulting from *k*-clique community detection with *k* values ranging from 5 to 12 are shown in Figure 2. There are many small communities distributed throughout the network (above 250 when *k* =3), and the number goes to 0 quickly (*k* = 11). This finding indicates that there are not many communities of tightly connected individuals, suggesting that the network does not take the form of a grassroots campaign in which many users produce and share information with one another (e.g., Bandari, Zhou, Qian, Tangherlini, & Roychowhur, 2017).

The lack of large, tightly connected clusters further suggests that the early Q network may have been easily disrupted had influential accounts been taken down. At a much later stage in QAnon development (January 2021), the removal of Trump and other prominent Q accounts led to a reduction of misinformation by 73% (Dwoskin & Timberg, 2021), highlighting the integral role of users central to the Q conspiracy on Twitter.

## Study 2: Modeling Tweet Content

We examined the semantic themes shaping discussion in the echochamber by applying the topic modeling algorithm *latent Dirichlet allocation* (LDA; Blei, Ng, & Jordan, 2014) to the tweets of the network's most central users. We chose to analyze the tweets of users who are highly central to the network, because many of the accounts in the network spread rather than produced information. Furthermore, a cursory analysis of the tweets from the users more peripheral to the network suggests that many of these users were bots and not ordinary human users. These users shared rather than produced original content, which is a hallmark of bot behavior (e.g., Lazer et al., 2018), and the limited content that they produced was largely incoherent. We aimed to limit our analyses to tweets from human users, such as those central to the production of content revealed in Study 1.

Specifically, we extracted the semantic content tied to users in different communities returned by calculating cliques of size 9. This selection returned a set of five distinct communities, allowing us to qualitatively assess whether there are clear differences in the semantic content in the tweets produced by users of each of the five communities. We considered this number to be a "goldilocks" value for qualitatively exploring semantic content present in the dataset (larger values of *k* produce too few communities, whereas smaller values of *k* produce too many distinct communities). Thus, we fit five topic models to the hashtags in tweets of five clusters of users uncovered in Study 1 to examine the extent to which semantic themes varied across the network.

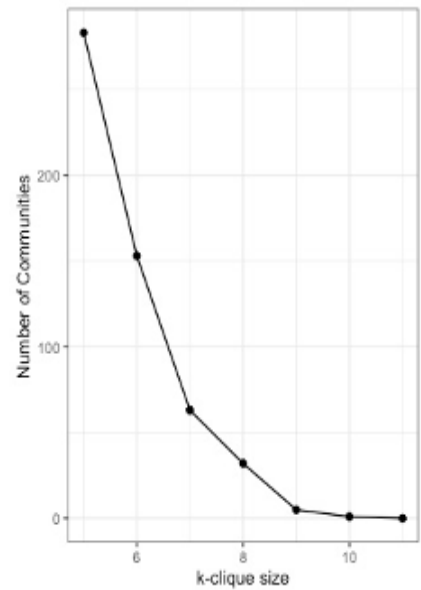

*Figure 2.* Number of communities uncovered via the *k*-clique community detection across values of *k*.

### Building Community-Specific Topic Models

LDA assigns each document (hashtags in a tweet) to topics by instantiating a set of *n* topics, where each topic is defined by a set of words that compose the documents. The algorithm assigns a probability value for how likely a document belongs to each topic. This probability value is determined by how many of the topic's "representative" words appear in the tweet. Analysis of the keywords describing the topics can shed light on the broad semantic themes being discussed in a corpus.

Table 2 lists a subset of topics and their associated keywords returned by the five community-specific topic models. Common clusters of topics are prevalent across all five communities. The semantic themes of each topic suggest that people are framing their discussion as a battle between *Forces for Good* as well as both *Unobservable* and *Observable Maleficent Forces.* This interpretation is supported by recent journalism work on Q believers (e.g., LeFrance, 2020). Topics representative of *Forces for Good* include: Guns (suggesting their militarization); Trump; Rainmakers; Q, QAnon; "Where we go one, we go all", and Michael Flynn. *Unobservable Maleficent Forces* included: DeepState; Human Trafficking Network; whereas *Observable Maleficent Forces* included: Mainstream Media and DemocRats. The theme of militarization is pervasive, revealed not only as a topic solely devoted to Guns in cluster 1, but also in the consistent presence across clusters of terms such as: LetTheTribunnalsBegin, PeopleAreWeapons, and PatriotsFight. Early Q believers apparently envisioned a literal battle between good and evil forces, foreshadowing the role played by Q supporters in the violent insurrection that occurred in January 2021 in Washington, D.C.

*Table 2*. Community-specific topics and keywords.

| Cluster | Topics | Keywords |
|---|---|---|
| 1 | Guns | Guns, GunsAreTools, PeopleAreWeapons |
| | Deep State | DeepStateCabal, SuperElite, Illuminati |
| | Gun Control | Socialists, Guns, GunGrabbers, SuperElite |
| 2 | Media | MSM (Mainstream Media), QRevolution, FakeNews |
| | Veteran | GodBlessOurTroops, PatriotsFight, MAGAveteran |
| 3 | Deep State | LetTheTribunalsBegin, Rainmakers, DeepState |
| 4 | Conspiracy | FakeNews, PerkinsCoie, PedoGate, PizzaGate |
| | Trafficking | DrainTheSwamp, SaveTheChildren, HumanTrafficking |
| 5 | Trafficking | HumanTrafficking, PerkinsCoie, ChildTrafficking |
| | MAGA | MAGA, Trump, POTUS, TheStorm |

*Note*. Not all topics from topic models reported here.

A general schema of Good versus Evil has long been exploited to engage public support for political issues (e.g., Lakoff, 1991). Understanding how this schema shapes reasoning in QAnon can help better understand how the narrative developed. Further, extensive research has found individual differences in susceptibility to conspiratorial thinking (Swami et al., 2011). Naturalistic work on factors leading to people's early involvement in Q can help understand which individual differences create vulnerability to conspiracy beliefs. The most common terms in user descriptions of the profiles in this dataset shed preliminary light on these factors. As shown in Table 3, users often self-identify as patriots and Christians. Given this exploratory finding, we propose that one's social identification (e.g., Christian, Patriot) leads to an adoption of a belief heuristic (Good vs Evil) that is used to select a specific political stance (e.g., gun rights) (see Table 4).

*Table 3*. Most frequent terms in user descriptions.

| Term | Proportion | | |
|---|---|---|---|
| MAGA | .27 | patriot | .07 |
| Trump | .19 | Christian | .07 |
| love | .11 | proud | .07 |
| conservative | .11 | country | .06 |
| God | .09 | NRA | .06 |

*Table 4*. Proposed social-cognitive model of QAnon belief formation.

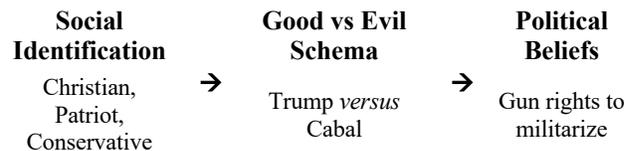

| Social Identification | Good vs Evil Schema | Political Beliefs |
|---|---|---|
| Christian, Patriot, Conservative | → Trump *versus* Cabal → | Gun rights to militarize |

## Discussion

Analysis of the QAnon Twitter network revealed that the majority of users shared, rather than produced, information. This finding suggests that development of the QAnon narrative, and its effects on shaping the beliefs of those in the network, were driven by a few key users. We applied topic modeling to analyze the tweet content of these users. This analysis revealed that users may hold a simplistic mental model in which political events are generated from an antithetical struggle between Evil (both observable and unobservable) and Good forces.

Our hypothesis about the mental schema of a QAnon user is currently based on an exploratory analysis of a naturalistic dataset. Future work—both naturalistic and empirical—is therefore required. More detailed topic modeling (e.g., models fit to the full text bodies of tweets) can shed further light on the semantic structure of this tweet corpus. A study that examines how endorsement of conspiratorial narratives changes as a function of their framing (e.g., as a struggle of good versus evil as compared to alternative framings) could shed light on the causal connection between this mental schema and propensity to engage in conspiratorial thinking. Additional demographic data could be used to assess which forms of social identification (e.g., religion, political beliefs) correlate most strongly with belief in a good-versus-evil archetype, and with belief in conspiracies more generally. It is possible that certain individual differences could predict which people are especially prone to fall for conspiracies when they are framed as a struggle between good versus evil. Such additional studies could help to elaborate and test the social-cognitive model we have proposed here.

## Acknowledgements
Preparation of this paper was supported by NSF Grant BCS-1827374 to KH.